\documentclass[sn-mathphys-num,Numbered]{sn-jnl}

\usepackage{graphicx}%
\usepackage{multirow}%
\usepackage{amsmath,amssymb,amsfonts}%
\usepackage{amsthm}%
\usepackage{mathrsfs}%
\usepackage[title]{appendix}%
\usepackage{xcolor}%
\usepackage{textcomp}%
\usepackage{manyfoot}%
\usepackage{booktabs}%
\usepackage{algorithm}%
\usepackage{algorithmicx}%
\usepackage{algpseudocode}%
\usepackage{listings}%
\usepackage{natbib}
\usepackage[varvw]{newtxmath}
\usepackage{newtxtext}
\usepackage{siunitx}[=v2]

\newcommand{\lapprox} {\, \lower3pt\hbox{$\sim$}\llap{\raise2pt\hbox{$<$}}\,}
\newcommand{\gapprox} {\, \lower3pt\hbox{$\sim$}\llap{\raise2pt\hbox{$>$}}\,}

\newcommand{\aap}{A\&A}

\graphicspath{{./}{figures/}}

\begin{document}

\title[Article Title]{Solar flares as electron accelerators: toward a resolution of the acceleration efficiency issue}


\author[1]{\fnm{Anna} \sur{Volpara}}

\author[2]{\fnm{Paolo} \sur{Massa}}

\author[1,3]{\fnm{Michele} \sur{Piana}}

\author[1]{\fnm{Anna Maria} \sur{Massone}}

\author[4]{\fnm{A. Gordon} \sur{Emslie}}

\affil[1]{\orgdiv{MIDA, Dipartimento di Matematica}, \orgname{Università di Genova}, \orgaddress{\street{via Dodecaneso 35}, \city{Genova}, \postcode{16145}, \country{Italy}}}

\affil[2]{\orgdiv{Institute for Data Science}, \orgname{University of Applied Sciences and Arts Northwestern Switzerland}, \orgaddress{\street{Bahnhofstrasse 6}, \city{Windisch}, \postcode{5210}, \country{Switzerland}}}

\affil[3]{\orgdiv{Osservatorio Astrofisico di Torino}, \orgname{Istituto Nazionale di Astrofisica}, \orgaddress{\street{Via Osservatorio 20}, \city{Pino Torinese}, \postcode{10025}, \country{Italy}}}

\affil[4]{\orgdiv{Department of Physics \& Astronomy}, \orgname{Western Kentucky University}, \orgaddress{1906 College Heights Boulevard}, \city{Bowling Green}, \state{KY}, \postcode{42101}, \country{USA}}


\abstract{A major open issue concerning the active Sun is the effectiveness with which magnetic reconnection accelerates electrons in flares. A paper published by {\em{Nature}} in 2022 used microwave observations to conclude that the Sun is an almost ideal accelerator, energizing nearly all electrons within a coronal volume to nonthermal energies. Shortly thereafter, a paper published in {\em{Astrophysical Journal Letters}} used hard X-ray measurements \emph{of the same event} to reach the contradictory conclusion that less than 1\% of the available electrons were accelerated. Here we address this controversy by using spatially resolved observations of hard X-ray emission and a spectral inversion method to determine the evolution of the electron spectrum throughout the flare. So we estimated the density of the medium where electrons accelerate and, from this, the ratio of accelerated to ambient electron densities. Results show that this ratio never exceeds a percent or so in the cases analyzed.}

%
%
%




\maketitle

\section{Main}\label{sec:main}

The standard ``thick-target'' model \citep{1971SoPh...18..489B,1973SoPh...31..143B,1976SoPh...50..153L} of solar flares involves the acceleration of electrons high in the corona by electric fields associated with magnetic reconnection; the electrons subsequently propagate down to the thicker regions of the chromosphere where they eventually thermalize. Bremsstrahlung collisions of the accelerated electrons, predominantly on ambient ions, lead to the emission of hard X-rays, and synchrotron radiation is produced as the accelerated electrons spiral around the magnetic field lines. Coulomb collisions, predominantly electron-electron, lead to the transfer of the accelerated electron energy into heat, resulting in both increased temperatures and pressure enhancements that drive dense chromospheric material upward into the coronal portion of the loop. The combined effects of increased temperature and increased coronal density produce the observed enhancements in thermal X-ray emission from extended coronal loops. The observed properties of hard X-ray, microwave, and soft X-ray radiation (see, e.g., \cite{2011SSRv..159...19F}) lend strong support to this model and thus establish that the energy release mechanism that triggers a solar flare involves a powerful electron accelerator.

One approach to computing the effectiveness of the electron acceleration process involves evaluating the ratio between two quantities: the average\footnote{Both the accelerated number density $n_{\rm acc}$ and the target number density $n_{\rm target}$ below may vary both within the field of view and along the line of sight. The ``average'' quantities used herein reflect both.} number density ${\overline n}_{\rm acc}$ (cm$^{-3}$) of the accelerated electron population that produces hard-X-rays, microwaves and heating, and the average number density ${\overline n}_{\rm target}$ of the available electrons in the ambient plasma.
Two recent results have provided dramatically different estimates of this quantity. On the one hand, spatially resolved observations of thermal and non-thermal electrons in a solar flare, inferred from microwave observations \cite{2022Natur.606..674F}, reached the conclusion that a solar flare can accelerate nearly \emph{all} the electrons within a relatively large coronal volume; on the other hand, hard X-ray observations \emph{for the same event} \cite{2023ApJ...947L..13K} led to the contradictory conclusion that only a very small fraction ($\lapprox 1\%$) of the electrons in the region was accelerated.

Since hard X-rays are produced primarily by bremsstrahlung collisions of accelerated electrons on ambient protons (which, by quasi-neutrality, have the same number density as electrons) the hard X-ray flux from a given volume is proportional to the product ${\overline n}_{\rm acc} \times {\overline n}_{\rm target}$; without additional information, only this product (cf. equation~(7) of ref. \cite{2023ApJ...947L..13K}) can be deduced. In this study, we provide this additional information. Specifically, we used observed spatial properties of hard X-ray emission and a collisional model of electron transport, to deduce the target density ${\overline n}_{\rm target}$ independently, thus disentangling the product ${\overline n}_{\rm acc} \times {\overline{n}}_{\rm target}$ and providing the values of both ${\overline n}_{\rm acc}$ and ${\overline{n}}_{\rm target}$. From this follows a reliable estimate of the ratio ${\overline n}_{\rm acc}/{\overline n}_{\rm target}$ and hence the effectiveness of solar flares as electron accelerators.

Our approach involved three main elements:

\begin{enumerate}
    \item Hard X-ray observations of loop-shaped flares provided by the Spectrometer/Telescope for Imaging X-rays (STIX; \cite{krucker2020spectrometer}) onboard the ESA Solar Orbiter mission. Native STIX data represent spatial Fourier components of the emitted X-ray flux collected at fixed spatial frequency points.
    
    \item A computational approach \cite{2007ApJ...665..846P} to imaging spectroscopy that is based on inverse problem theory. This method produces maps of the accelerated electrons responsible for the X-ray emission that are smoothed over a range of electron energies. Accordingly, the electron spectra constructed from these maps are regularized, which allows straightforward and reliable inference of their evolution with position throughout the flare volume.
    
\item Finally, the application of a collisional electron transport model to the variation of the electron flux spectrum with position leads to an estimate of the average target density ${\overline n}_{\rm target}$. This independent evaluation of ${\overline n}_{\rm target}$ allows each of the terms in the product ${\overline n}_{\rm acc} \, \times \, {\overline n}_{\rm target}$, and hence their all-important ratio ${\overline n}_{\rm acc}/ {\overline{n}}_{\rm target}$, to be determined.

\end{enumerate}

Analysis of several flares observed by STIX showed that the quantity $\eta \equiv {\overline n}_{\rm acc}/{\overline n}_{\rm target}$ assumes values that depend on the flare involved, but that are not correlated with the intensity of the flare. Rather, $\eta$ decreases with the mean target density ${\overline n}_{\rm target}$, which we show to be an expected feature of models involving acceleration by relatively weak, large-scale, electric fields. This robust measure of the acceleration efficiency is typically around a few percent, and never exceeds $2\%$ for any of the events considered in this study.

\section{Results}

\subsection{STIX data}

\begin{figure}
\centering
\includegraphics[width=13.cm]{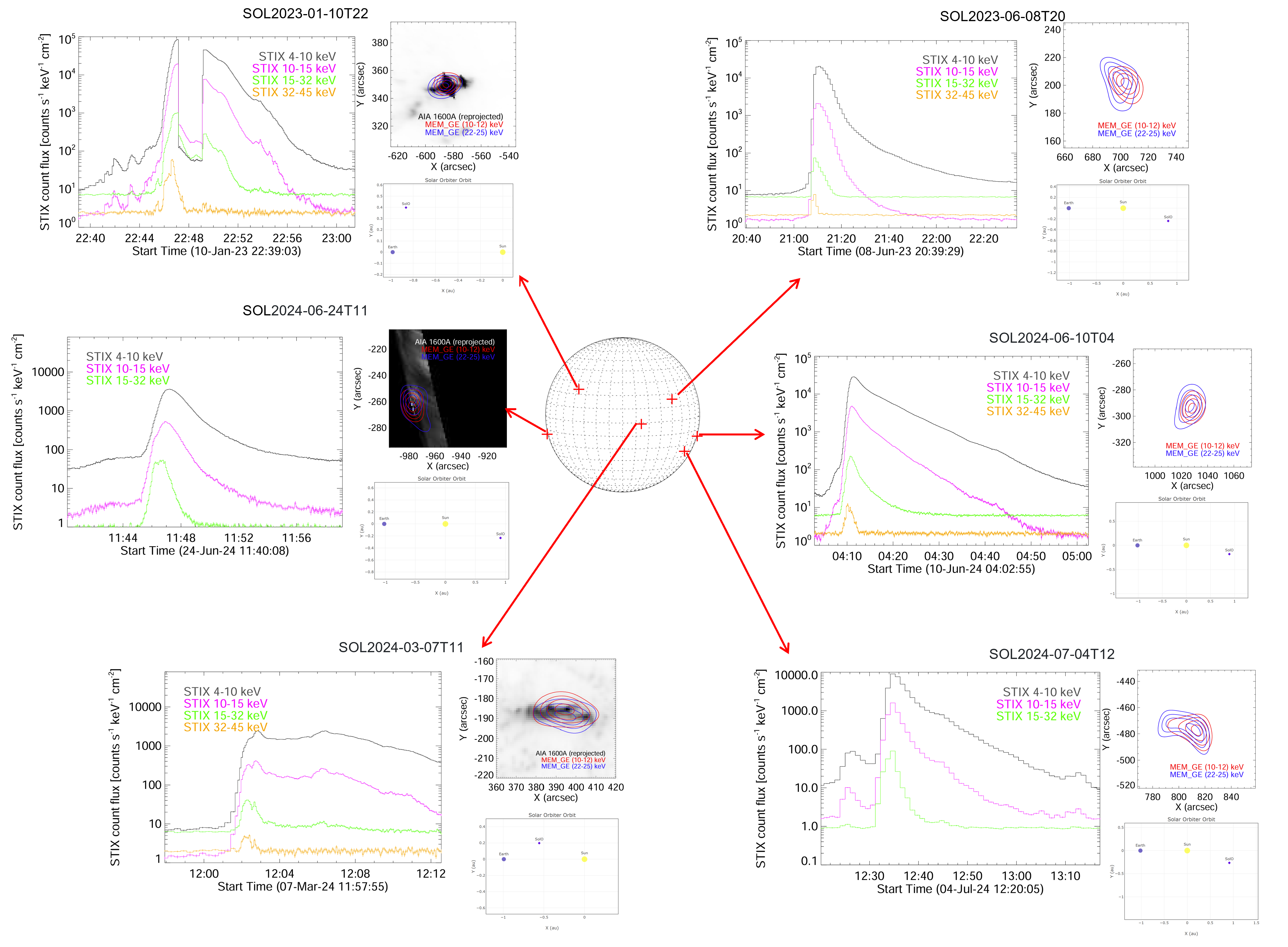}
\caption{The six flares considered in this study are shown from the reference field of view of Solar Orbiter at the time of each observation. For each event, the left panels show the STIX light-curves in the count energy bands shown; the top right panels show the hard X-ray flux contours obtained by applying the MEM$\_$GE image reconstruction method to the STIX 10-12~keV (thermal) data (red) and 22-25 keV (non-thermal) data (blue); the bottom right panels illustrate the positions of Solar Orbiter and the Earth with respect to the emission region. The three events on the left of the Figure were also observed by AIA, and the hard X-ray level contours have been superimposed onto the 1600~\AA\ EUV maps.}\label{fig:events}
\end{figure}

The approach used requires events, selected from the STIX database, that present a clearly visible loop shape, so that the variation of the electron spectrum along the magnetic field lines that define the loop axis can be discerned. Suitable events are also characterized by more than $3000$~counts at photon energies higher than $25$~keV, to provide a sufficiently high signal-to-noise ratio. Finally, the event should preferably be at a position on the Sun that is also visible from the Earth, in order to leverage morphological information from the Atmospheric Imaging Assembly (AIA; \cite{2012SoPh..275...17L}) onboard the Solar Dynamics Observatory (SDO). We selected the six flares shown in Figure~\ref{fig:events}, which illustrates the position on the Sun for each event, together with the corresponding STIX light-curves. STIX count level contours were obtained using the visibility-based MEM$\_$GE \cite{2020ApJ...894...46M} image reconstruction method. Three of these events (left side of Figure \ref{fig:events}) were also observed by AIA, and for these the hard X-ray contours have been superimposed onto the AIA $1600~\AA$ maps. For the other three events, Figure~\ref{fig:events} provides only the hard X-ray contour maps. Table~\ref{tab:events} provides selected details of the flares.



\begin{table}[t]
	\centering
	\scriptsize
	\resizebox{\columnwidth}{!}{
		\begin{tabular}{cccccc}
			\toprule
			Event & Time Interval (UT) & GOES Class & \multicolumn{2}{c}{Energy Range (keV)} \\
			\cmidrule(lr){4-5}
			& & & Thermal & Non-Thermal \\
			\midrule
			SOL$2023-01-10$T$22$ & 22:44:24 -- 22:46:24 & X$1.1$ & 10--12  & 22--25 \\
			SOL$2023-06-08$T$20$ & 21:07:00 -- 21:09:00 & C$1.5$ & 10--12 & 22--25 \\
			SOL$2024-03-07$T$11$ & 11:58:00 -- 12:00:00 & C$6.3$ & 10--12  & 22--25 \\
			SOL$2024-06-10$T$04$ & 04:08:34 -- 04:10:34 & C$5.9$ & 10--12  & 22--25 \\
			SOL$2024-06-24$T$11$ & 11:45:19 -- 11:47:19 & M$1.5$ & 10--12  & 22--25 \\
			SOL$2024-07-04$T$12$ & 12:34:00 -- 12:36:00 & C$4.5$ & 10--12  & 22--25 \\
			\bottomrule
		\end{tabular}}
	\caption{Parameters of the events considered in this study and illustrated in Figures \ref{fig:events}. For each event the table contains the time interval of STIX observations, the flare GOES class, and the two energy ranges used to construct the hard X-ray contours in Figure~\ref{fig:events}.}
	\label{tab:events}
\end{table}

\subsection{Spatially resolved electron spectra agree with the standard flare model}\label{sec:spectra_agree}

\begin{figure}
\centering
\includegraphics[width=13.cm]{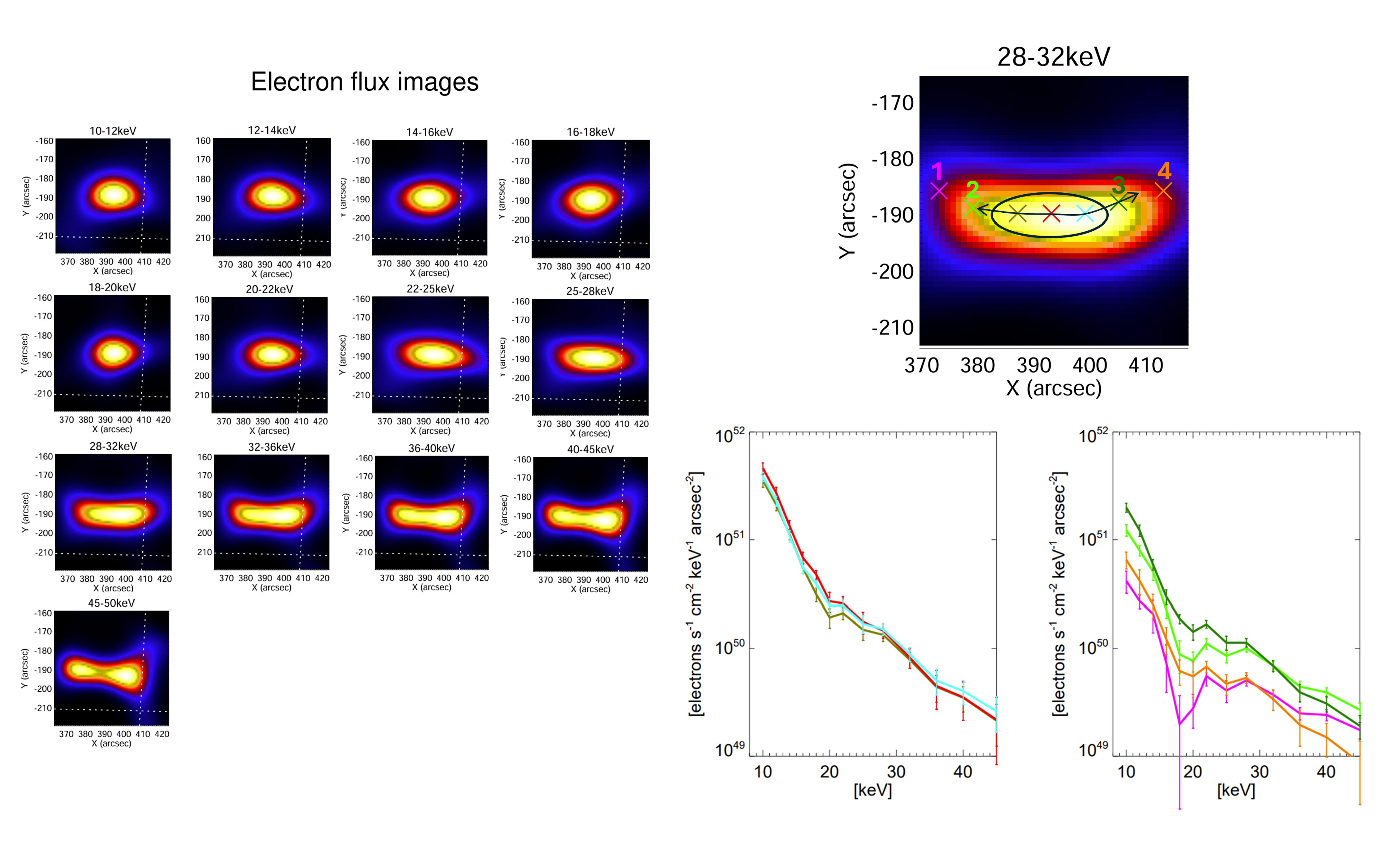}
\caption{Imaging spectroscopy analysis performed in the electron domain for the March 7 2024 flare. The left panel contains 13 electron flux maps reconstructed from STIX visibilities at 13 electron energy channels. The top right panel highlights seven positions in the loop in the 28-32 keV energy range. The bottom right panel shows electron flux spectra corresponding to the three crosses in the acceleration region (left panel) and the four positions where the Coulomb collisional model is applicable active.}\label{fig:electron-domain-single}
\end{figure}

We analyzed the STIX visibilities via an imaging spectroscopy method \citep{2007ApJ...665..846P} that exploits the linearity of both the inverse spatial Fourier transform and the photon $\rightarrow$ electron spectral inversion procedures to perform them in the reverse order: spectral, then spatial.  First, the count visibilities represented in the native STIX data are spectrally inverted, using the fully relativistic, isotropic, Bethe-Heitler BN bremsstrahlung cross-section \citep{1959RvMP...31..920K}, to obtain the corresponding visibilities of the electron flux. Then, MEM$\_$GE is applied in the \emph{electron} domain, using the electron flux visibilities as input. Since the initial spectral inversion process is performed by means of a regularization method, the resulting electron flux images necessarily vary smoothly with energy, and ``stacking'' these images thus allows a reliable determination of the electron flux spectrum at each location in the image. Figure~\ref{fig:electron-domain-single} provides an example of the results obtained by this approach, for the 2024~March~7 event. The pixel content of each map in the left panel of this figure corresponds to the quantity ${\cal{F}}(x,y;E)$, equal to the mean electron flux spectrum ${\overline F} (x,y;E)$ (electrons~cm$^{-2}$~s$^{-1}$~keV$^{-1}$) in a specific electron energy channel and in that pixel, weighted by the target column density ${\overline n}_{\rm target}(x,y) \, \ell (x,y)$ of the hard X-ray source along the line of sight, i.e.,

\begin{equation}\label{eq:pixel-content}
    \mathcal{F}(x,y;E) = a^2 \,\, {\overline n}_{\rm target}(x,y) \, \ell(x,y) \, {\overline F}(x,y; E) \,\,\, ,
\end{equation}
where $a$ is the conversion factor from arcsec to cm ($= 7.25 \times 10^7 \, R$, where $R$(AU) is the Sun-spacecraft distance).

We point out that the events studied here all have a spatial structure that exhibits a loop-like structure over a wide range of electron energies, as for the 2024 March~7 event (and as for the flare described in \citep{2024A&A...684A.185V}). Thus, for all events considered for this study, the density in the coronal source is sufficiently high that a substantial fraction of the hard X-rays are emitted from the corona, and the observed visibilities can accurately determine the coronal component of the hard X-ray emission. It follows that the electron maps produced from such data permit an evaluation of the shape of the electron flux spectrum in a direction along the projection of the loop on the plane of the sky.

In the 28-32~keV map at the top right panel of Figure~\ref{fig:electron-domain-single}, seven positions are shown, superimposed on the reconstructed electron flux distribution. The bottom panels show the electron flux spectra at each of these positions, where the colors of the spectra match the colors of the selected locations. The large central area in which the thermal component of the emission is dominant (see the locations denoted with the $X$ symbol in the panel) is taken to represent the electron acceleration site, while the other points (see the locations denoted with 1, 2, 3, and 4 in the panel) are in the ``target'' region through which the accelerated electrons subsequently propagate. As the distance from the acceleration site increases, the electron flux generally decreases in magnitude, and the local maximum in the non-thermal portion of the spectrum shifts to higher energies. Both these behaviors are consistent with a scenario in which the X-ray-emitting electrons are slowed by interaction with ambient electrons, with an energy loss rate that decreases with energy, so that there is a progressive depletion of the low energy part of the distribution, with only the most energetic electrons able to reach large distances from the acceleration site. 
This scenario is confirmed for all the events selected in this study.

\subsection{The accelerated electron fraction never exceeds a few percent}\label{acc_rate_percent}

The acceleration/target scenario illustrated in the previous subsection may be modeled collisionally. The injection of a power-law spectrum of electrons $F(s_o;E) = A E^{-\delta}$ at position $s_o$ into a target of average density ${\overline n}_{\rm target}$, in which the electrons lose energy at a rate given by the form appropriate to Coulomb collisions \citep{1978ApJ...224..241E}, viz. $dE/ds = - 4 \pi \, e^4 \, \ln \Lambda \, {\overline n}_{\rm target}/E$ (where $e$ is the electronic charge and $\ln \Lambda$ the Coulomb logarithm), results in a spatial variation of the electron spectrum throughout the target, with the form

\begin{equation}\label{eq:collisional-solution-main}
{\overline F}(s; E) = A \, \frac{E}{\left  ( E^2 + E_{\rm stop}^2(s-s_o) \right ) ^{(\delta+1)/2} } 
\end{equation}
where $E_{\rm stop} = \sqrt{4 \pi \, e^4 \, \ln \Lambda \, {\overline n}_{\rm target} \, (s - s_o)}$ is the energy that an electron must have at injection to (just) reach the point $s$. 
Consistent with the electron spectra inferred from the STIX hard X-ray data (Figure~\ref{fig:electron-domain-single}), such a spectrum exhibits a local maximum at an energy $E_{\rm max} = E_{\rm stop}/\sqrt{\delta}$, which increases like $\sqrt{{\overline n}_{\rm target} \, (s-s_o)}$.

\begin{figure}
\centering
\includegraphics[width=10.cm]{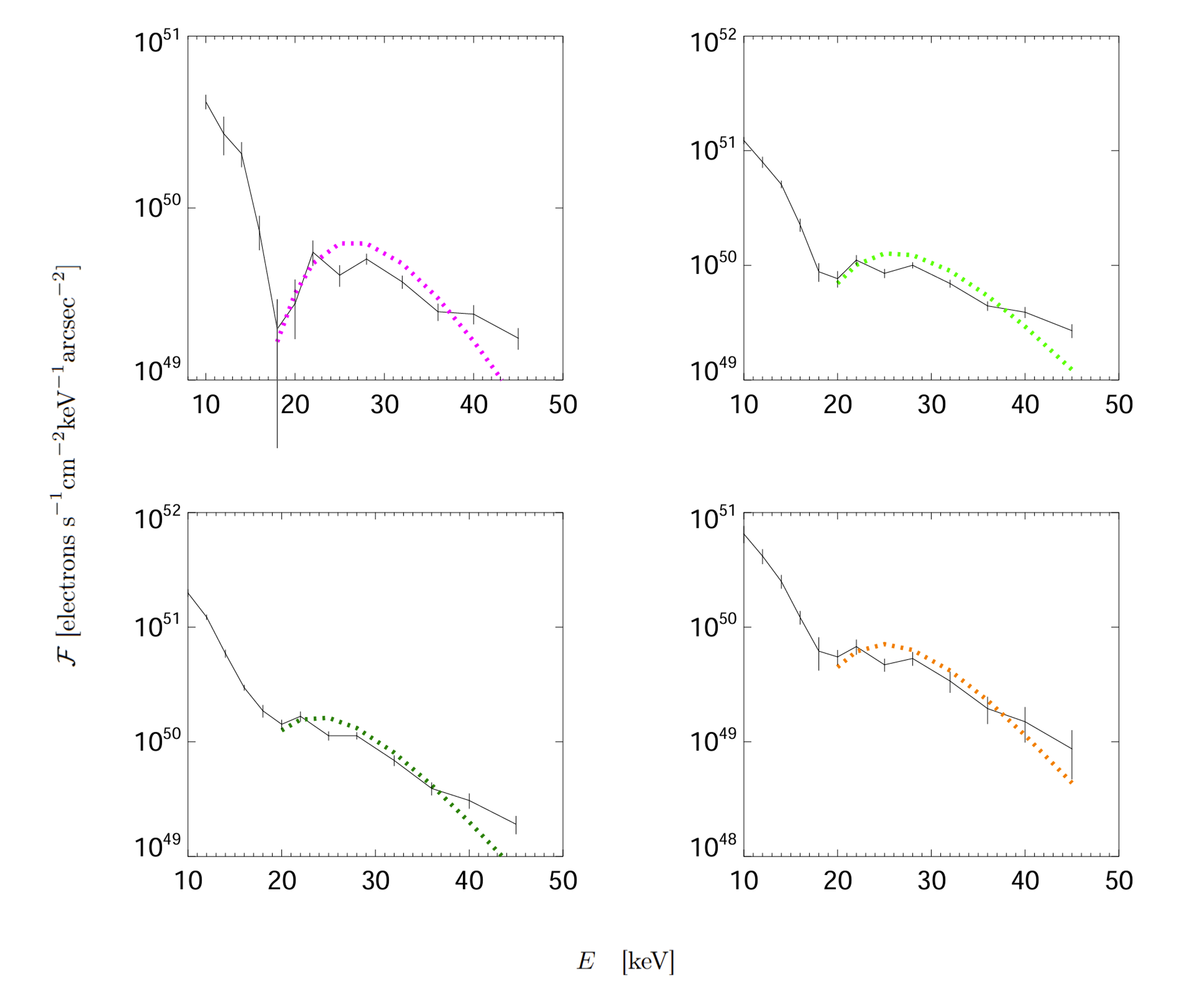}
\caption{Regression analysis of the spatially-resolved spectra at non-thermal energies. The data points represent the electron spectra for the March 7 2024 event at the selected points within the loop.  Fits to the non-thermal part of the spectrum, following equation~\eqref{eq:dense-acceleration-region-square-text}, with $L = 5.2 \times 10^8$~cm, are shown as dotted lines, with the color of the line matching the color used to define the points 1, 2, 3, and 4 in Figure~\ref{fig:electron-domain-single}. Note the strong thermal component at low energies; the data points at those energies were not used for the fit. }\label{fig:model-fit}
\end{figure}

Convolution of the spectrum in equation~\eqref{eq:collisional-solution-main} with a finite acceleration region of length $L$ gives a spectrum of the form

\begin{equation}\label{eq:dense-acceleration-region-square-text}
 F(s; E) = A \, \, \int_{s_o = -L/2}^{L/2}
  \, \frac{E}{\left  ( E^2 + 2 \, K \, {\overline n}_{\rm target} \vert s-s_o \vert \right ) ^{(\delta+1)/2} } \, ds_o \,\,\, .
\end{equation}
Fitting expression~\eqref{eq:dense-acceleration-region-square-text} to the shapes of the forms of the spectra ${\cal F} (s;E)$ inferred from the STIX data (Figure~~\ref {fig:model-fit}) thus allows a determination of the mean target density ${\overline n}_{\rm target}$, and hence, using Equation~(\ref{eq:pixel-content}), the local electron flux spectrum $F (s; E)$. 
If we further introduce the \textit{mean accelerated electron number density spectrum} (electrons~cm$^{-3}$~keV$^{-1}$):

\begin{equation}\label{eq:n_acc_def}
\frac{d{\overline n}_{\rm acc}}{dE} (x,y;E) = \, \frac{{\overline F}(x, y; E)}{v(E)}= \sqrt{\frac{m_e}{2 \, E}} \,\, {\overline F}(x, y; E)  ~,
\end{equation}
it follows from equations~\eqref{eq:pixel-content} and~\eqref{eq:n_acc_def} that

\begin{equation}\label{nacc-n-electron-map}
\frac{d{\overline n}_{\rm acc}}{dE} (x,y; E) \,\, {\overline n}_{\rm target}(x,y) = \sqrt{\frac{m}{2E}} \, \frac{\mathcal{F}(x,y;E)}{a^2 \, \ell(x,y)}  = 
\frac{5.3\,  \times \, 10^{-10}}{\sqrt{E \, {\rm (keV)}}} \, \frac{\mathcal{F}(x,y;E)}{a^2 \, \ell(x,y)} \, \,\,\, .
\end{equation}
With the mean ambient target density ${\overline n}_{\rm target}$ already determined from analysis of the electron spectrum evolution with position, as described above, the accelerated electron number spectrum $d{\overline n}_{\rm acc}/dE (x, y; E)$ can be straightforwardly calculated from the STIX data. (Since we have no direct information on $\ell$, the extent of the source along the line of sight, we assumed a roughly circular loop cross-section and simply take $\ell$ to be the same as the width of the loop structure on the plane of the sky.) The accelerated number density (cm$^{-3}$) at position $s(x,y)$ along the loop is then found from ${\overline n}_{\rm acc} (s) = \int (d{\overline n}_{\rm acc}/dE) (s; E) \, dE $. The integration limits were taken to be from a low-energy cutoff of $20$~keV, with an upper limit at $45$~keV, because of the unreliable nature of the inferred electron spectrum at higher energies. 

\begin{table}[t]
	\centering
	\scriptsize
	\resizebox{\columnwidth}{!}{
		\begin{tabular}{c|ccc|ccc}
			\toprule
            & \multicolumn{3}{c}{Acceleration Region Parameters} & \multicolumn{3}{|c}{Acceleration Efficiency} \\
            \cmidrule{2-7}
			Event & $L$ (cm) & $\ell $ (cm) & $T$ (K) & ${\overline n}_{\rm target}$ (cm$^{-3}$) & ${\overline n}_{\rm acc}$ (cm$^{-3}$) & $\eta$ \\
			\midrule
			SOL$2023-01-10$T$22$ & $7.1 \times 10^8$ & {$5.0 \times 10^{8}$} & $2.2 \times 10^7$ & $1.2 \times 10^{10}$ &  $1.8 \times 10^8$ & 0.015  \\
			SOL$2023-06-08$T$20$ & $4.4 \times 10^8$ & {$1.0 \times 10^{9}$}  & $1.5 \times 10^7$ & $1.1 \times 10^{10}$ & $2.5 \times 10^7$ & 0.002 \\
			SOL$2024-03-07$T$11$ & $5.2 \times 10^8$ & {$5.0 \times 10^{8}$} & $1.7 \times 10^7$ & $1.6 \times 10^{10}$ & $1.9 \times 10^7$ & 0.001 \\
			SOL$2024-06-10$T$04$ & $1.1 \times 10^9$ & {$9.0 \times 10^{8}$} & $2.0 \times 10^7$  & $8.5 \times 10^9$ & $7.6 \times 10^7$ & 0.009 \\
			SOL$2024-06-24$T$11$ & $4.0 \times 10^8$ & {$6.0 \times 10^{8}$} & $1.4 \times 10^7$ & $3.6 \times 10^{10}$ & $5.0 \times 10^7$ & 0.001 \\
			SOL$2024-07-04$T$12$ & $1.0 \times 10^9$ & {$8.0 \times 10^{8}$} & $1.8 \times 10^7$ & $1.9 \times 10^{10}$ & $4.5 \times 10^7$ & 0.004 \\
			\bottomrule
		\end{tabular}}
	\caption{Model parameters for each event. The first four columns contain the date of the event and the length, width, and temperature of the homogeneous region at the top of the loop structure, identified as the acceleration region. The final three columns give the mean target density and accelerated number density, deduced from fitting the spatial variation of the electron spectrum, and finally the inferred value of the ratio $\eta = n_{\rm acc}/{\overline n}_{\rm target}$.}
	\label{tab:parameters}
\end{table}

With the values of ${\overline n}_{\rm acc}$ and ${\overline n}_{\rm target}$ thus determined, we can now obtain the value of the ratio $\eta = {\overline n}_{\rm acc} \, / \, {\overline n}_{\rm target}$. This is a measure of the efficiency with which the primary energy release mechanism accelerates electrons to deka-keV nonthermal energies, and its value imposes very significant constraints on the mechanism responsible for accelerating the hard-X-ray-producing electrons.

Table~\ref{tab:parameters} shows the results obtained. In each row we give the date of the event, and then in the next three columns the properties of the acceleration region, identified as the region of roughly homogeneous (thermal) X-ray intensity near the apex of the loop (see Figure~\ref{fig:electron-domain-single}):

\begin{itemize}
    \item $L$ (cm), its length, converted from arcseconds on the plane of the sky using the distance $D$ (AU) between Solar Orbiter and the Sun at the time of the event;
    \item $\ell$ (cm), its transverse width; and 
    \item $T$ (K), its temperature, deduced by fitting the STIX photon spectrum for that region with a form appropriate to an isothermal source.    
\end{itemize}
The last three columns contain first the value of ${\overline n}_{\rm target}$ (cm$^{-3}$), obtained from fitting the electron spectra in the remaining parts of the source, as a group, to the cold target collisional model (see subsection 4.4 in the ``Methods" section for details). Then, we give the accelerated number density ${\overline n}_{\rm acc}$, obtained from the intensity of the hard X-ray flux observed by STIX, using equation~\eqref{nacc-n-electron-map} with the value of $\ell$ from column~3. Finally, the last column contains the acceleration efficiency $\eta = {\overline n}_{\rm acc}/\overline{n}_{\rm target}$.

\subsection{Acceleration rate decreases with target density}\label{sec:interpretation}

Results in Table~\ref{tab:parameters} can be interpreted by means of a wide range of possible acceleration mechanisms \citep{1994ApJ...435..469B,2000SoPh..194..327L,1995A&A...304..576M,1996ApJ...461..445M,2011SSRv..159..357Z}. However, since at a fundamental level \emph{all} electron acceleration occurs under the action of a local field ${\cal E}$ in the electron's frame, we choose here to characterize the acceleration mechanism through the value of this fundamental accelerating electric field ${\cal E}$.

We shall therefore interpret the results of Table~\ref{tab:parameters} in the context of a simple acceleration model that invokes an accelerating electric field ${\cal E}$ and a collisional drag force characterized by a collision frequency with the $1/v^3$ velocity dependence appropriate to Coulomb collisions \cite{1962pfig.book.....S}. As shown in Section~\ref{sec:acceleration-model}, in such a scenario only electrons with a velocity  $v > v_{\rm crit} = \sqrt{v_{\rm th} \, ({\cal E}_D/{\cal E} )}$ are accelerated, where $v_{\rm th}$ is the thermal speed $= \sqrt{2 k_B T/m_e}$ (where $k_B$ is Boltzmann's constant, $T$ the ambient temperature and $m_e$ the electron mass), and ${\cal E}_D$ is the Dreicer \cite{1959PhRv..115..238D} field:

\begin{equation}\label{eq:Dreicer Field-alt}
    {\cal E}_D  \simeq 4 \times 10^{-5} \, \frac{({\overline n}_{\rm target}/10^{10} \, {\rm cm}^{-3})}{(T/10^7 \, {\rm K)}} \, \, {\rm V \, cm}^{-1}\,\,\, .
\end{equation}
The fraction $\eta$ of the ambient population that is accelerated is obtained by integrating a Maxwellian distribution over velocities $v > v_{\rm crit}$, so that (see Section~\ref{sec:acceleration-model}) for a given electric field ${\cal E}$,

\begin{equation}\label{eq:theoretical-efficiency}
\eta = \frac{1}{2} \, {\rm erfc} \left ( \left [ \frac{{\cal E}_D}{{\cal E}} \right ]^{1/2} \right ) \simeq \frac{1}{2} \, {\rm erfc} \left ( \left [ \frac{{\overline n}_{\rm target}}{n_o}  \right ]^{1/2} \right ) \,\,\, .
\end{equation}
Here ${\rm erfc}(z)$ is the complementary error function and we have assumed that the density in the acceleration region is comparable to the density ${\overline n}_{\rm target}$ in the remainder of the loop structure. The constant parameter\footnote{Physically, $n_o$ is the density for which the collisional drag force on an electron moving at the thermal speed (corresponding to temperature $T$) is equal to the force associated with the applied electric field ${\cal E}$, so that the applied field is equal to the Dreicer field ${\cal E}_D$ for the ambient medium.} $n_o = k_B T \, {\cal E}/ 2 \pi e^3 \ln \Lambda \simeq 2.5 \times 10^9 \, {\cal E}_{-5} \, T_7$, with ${\cal E}_{-5}$ the electric field in units of $10^{-5}$~V~cm$^{-1}$ and $T_7$ the temperature in units of $10^7$~K.

\begin{figure}
    \centering
    \includegraphics[width=1.\linewidth]{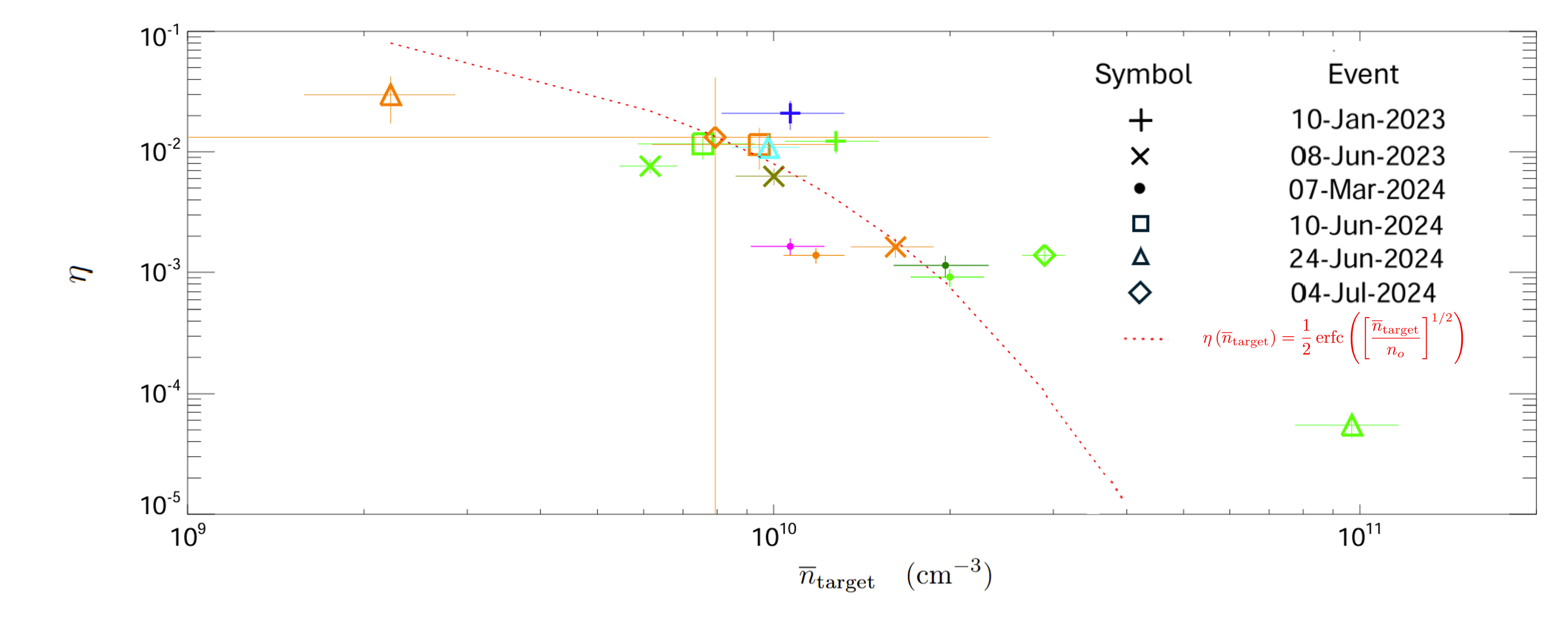}
    \caption{Scatter plots of the acceleration efficiency parameter $\eta$ versus $\overline{n}_{\text{target}}$, the pixel-averaged electron number density in the target. Each symbol is related to a particular event and each color refers to a different position within that event. The curve shows the fit to equation~\eqref{eq:theoretical-efficiency}, with $n_o = 5 \times 10^9$~cm$^{-3}$.}
    \label{fig:dreicer}
\end{figure}

Figure~\ref{fig:dreicer} shows a scatterplot of $\eta$ versus ${\overline n}_{\rm target}$ for the flares studied, confirming the expected generally decreasing trend of $\eta$ with ${\overline n}_{\rm target}$. This result is entirely plausible: from an observational viewpoint, a larger target density requires a smaller density of accelerated electrons to produce a given hard X-ray intensity, while from a theoretical viewpoint, a large target density increases the collisional drag force, so that a given applied field will accelerate a smaller fraction of the ambient population.  Fitting Equation~(\ref{eq:theoretical-efficiency}) to the plotted points gives $n_o \simeq 5 \times 10^9$~cm$^{-3}$, corresponding, for a temperature $T \simeq 10^7$~K, to an electric field strength ${\cal E} \simeq 2 \times 10^{-5}$~V~cm$^{-1}$.

\section{Discussion}\label{sec:discussion}

Although in this study we have considered only six events, there is still a large enough range of parameter values in Table~\ref{tab:parameters} to discuss the meaning and reliability of the results described in the previous section. The variation of $\eta$ with ${\overline n}_{\rm target}$ predicted by equation~\eqref{eq:theoretical-efficiency} (Figure~\ref{fig:dreicer}) allows us to construct a self-consistent scenario for the acceleration of electrons in the solar flares studied. An electric field of $\sim$$2 \times 10^{-5}$~V~cm$^{-1}$ operates in a medium of density $\sim$$1.5 \times 10^{10}$~cm$^{-3}$ and temperature $\sim$$10^7$~K, corresponding (equation~\eqref{eq:Dreicer Field-alt}) to a Dreicer field $\sim$$6 \times 10^{-5}$~V~cm$^{-1}$. Acting over an acceleration length $L \gapprox 10^9$~cm, such an electric field accelerates electrons to $\gapprox 20$~keV, consistent with the energies of hard-X-ray-producing electrons. The applied electric field is about a third of the Dreicer field, so that it accelerates $\sim$$0.5$\% of the ambient population (equation~\eqref{eq:theoretical-efficiency}), giving an accelerated particle density $\sim$$10^{8}$~cm$^{-3}$. Such a density, injected into a target of density $1.5 \times 10^{10}$~cm$^{-3}$, produces a hard X-ray flux that is consistent with observations (equation~\eqref{nacc-n-electron-map}).

However, the most significant result of this study is that in all the events studied the value of $\eta$ lies in the range $0.001 \lapprox \eta \lapprox 0.015$: the acceleration efficiency is in general less than a percent, and only a fraction of a percent in some cases, consistent with the conclusions in \cite{2023ApJ...947L..13K}. Given the importance of this result, and the significant difference between it and the claims in \cite{2022Natur.606..674F} of an efficiency $\eta \simeq 1$, it is worth commenting on the robustness of our conclusions.
To that end, we would point out that the treatment here requires a bare minimum of modeling assumptions.

First, the derived electron spectra follow straightforwardly from the hard X-ray visibilities observed by STIX, requiring only knowledge of the bremsstrahlung cross-section used in the spectral inversion. For the non-relativistic photon and electron energies considered here, possible deviations from the isotropic cross-section used have a negligible effect on the inferred electron spectrum (cf. Figure~4 of \cite{2004ApJ...613.1233M}).

Second, the inferred target density is determined by applying an energy loss model for the accelerated electrons that is characterized by the most basic (and unavoidable) of processes, namely Coulomb collisions on ambient electrons.

Finally, given a target density, determination of the accelerated number density requires only knowledge of the emitted hard X-ray flux from the region into which the electrons are injected, which is well determined from the STIX hard X-ray data.

In contrast, the conclusions in \cite{2022Natur.606..674F} involve interpretation of microwave observations, which requires a number of assumptions about, for example, the magnetic field configuration in the emitting region and the angular distribution of the accelerated electrons, both of which are generally unknown. Further, as noted in \cite{2023ApJ...947L..13K}, microwave emission is produced primarily by mildly relativistic electrons, so that determination of the overall accelerated electron number requires extrapolation of the electron spectrum over a considerable energy range. Hard X-ray observations indicate that the steep spectra commonly inferred from microwave observations do not continue down to the lower energies where the bulk of the accelerated electron population resides, and such a spectral flattening at low energies would greatly reduce the inferred density in acceleration electrons. It follows that there is room for far greater uncertainty in the result derived in \cite{2022Natur.606..674F}.

We conclude that while solar flares clearly do accelerate electrons, they do so with only modest efficiency, with \emph{less than a few percent of the ambient population undergoing acceleration to nonthermal energies}.


\section{Methods}

\subsection{The STIX imaging concept}

The Spectrometer/Telescope for Imaging X-rays (STIX) \cite{krucker2020spectrometer} is a hard X-ray imaging instrument mounted on the Solar Orbiter cluster, launched by the ESA in February 2020. 
The main scientific goal of STIX is to determine the intensity, spectrum, timing and location of accelerated electrons during solar flares. 
STIX hardware consists of 30~detectors recording X-ray photons in the range 4 - 150 keV. 
The X-ray flux incident on each detector is modulated by means of a bi-grid collimator, the configuration of which creates the superposition of two spatial modulations, named a Moir\'e pattern, with each pattern characterized by a specific vector in the two-dimensional spatial frequency domain \cite{giordano2015process,massa2019count,massa2023stix}. The intensity and phase of each Moir\'e pattern provides amplitude and phase information on the corresponding spatial Fourier component of the incoming flux (termed, following convention in the radio astronomy domain, a ``visibility''). The 30~visibilities measured by STIX can be used to reconstruct an image of the flaring X-ray emission using a regularized discrete Fourier transform inversion, in practice carried out using an image reconstruction method that is tailored to the sparse nature of the available information in the spatial frequency domain \cite{siarkowski2020non,2020ApJ...894...46M,perracchione2021visibility,volpara2022forward,perracchione2023unbiased,volpara2024multi}.



\subsection{Electron maps}

We define the {\it mean source electron spectral flux map} ${\overline F}(x,y; E)$ (electrons~cm$^{-2}$~s$^{-1}$~keV$^{-1}$) by performing a density-weighted average of the electron flux along the line of sight $z$ (cm):

\begin{eqnarray}\label{eq:fbar-definition}
{\overline F}(x,y;E) &=& \frac{\int_{z=0}^{\ell(x,y)} n_{\rm target}(x,y,z) \, F(x,y,z;E) \, dz} {\int_{z=0}^{\ell(x,y)} n_{\rm target}(x,y,z) \, dz} \cr &\equiv& \frac{1}{{\overline n}_{\rm target}(x,y) \, \ell(x,y)} \, \int_{z=0}^{\ell(x,y)} n_{\rm target}(x,y,z) \, F(x,y,z;E) \, dz \,\,\, ,
\end{eqnarray}
where $(x,y)$ (each in arcsec units) is a point in the image plane, $\ell(x,y)$ (cm) is the extent of the source along the line of sight, and $n_{\rm target}$ and $F(x,y,z;E)$ are the local density (cm$^{-3}$) and the electron flux spectrum (electrons~cm$^{-2}$~s$^{-1}$~keV$^{-1}$), at point $(x,y,z)$ in the source. Since the source is optically thin at X-ray energies, the corresponding \textit{photon spectral map} $I(x,y;\epsilon)$ (photons~cm$^{-2}$~s$^{-1}$~keV$^{-1}$~arcsec$^{-2}$) is simply \cite{2003ApJ...595L.115B}

\begin{eqnarray}\label{eq:fundamental}
I(x,y; \epsilon) &=& \frac{a^2}{4 \pi R^2} \, \int^{\infty}_{E =\epsilon}  \int_{z=0}^{\ell(x,y)} n_{\rm target}(x,y,z) \, F(x,y,z; E) \, Q(\epsilon, E) \, dE \, dz  \cr
&=& \frac{a^2}{4 \pi R^2} \, {\overline n}_{\rm target}(x,y) \, \ell (x,y) \, \int^{\infty}_{E =\epsilon} {\overline F}(x,y; E) \, Q(\epsilon, E) \, dE \,\,\, ,
\end{eqnarray}
where $R$ (cm) is the distance from the source to the instrument, $Q(\epsilon, E)$ (cm$^2$~keV$^{-1}$) is the cross-section\footnote{The form of the quantity $Q(\epsilon, E)$ depends on the emission process being considered. In principle, this could include a host of emission processes, such as gyrosynchrotron emission, inverse Compton emission, free-bound emission, and bremsstrahlung.  We here take the form of $Q(\epsilon, E)$ as that corresponding to bremsstrahlung, and we use the isotropic form of the cross-section in \cite{1959RvMP...31..920K}.} for photon emission, differential in photon energy $\epsilon$, and $a = R/206265 = 1.5 \times 10^{13} {\widetilde R}/206265 = 7.25 \times 10^7 \, {\widetilde R}$~cm~arcsec$^{-1}$ is the conversion factor from arcseconds to cm, ${\widetilde R}$ being the distance from the source to the observer in astronomical units (AU). Using equation~(\ref{eq:pixel-content}) in equation~(\ref{eq:fundamental}) leads to the following, more compact form, for $I(x,y;\epsilon)$:

\begin{equation}\label{eq:fundamental-1}
I(x,y; \epsilon) = \frac{1}{4 \pi R^2} \, \int^{\infty}_{E =\epsilon} \mathcal{F}(x,y;E) \, Q(\epsilon, E) \, dE \,\,\, .
\end{equation}


The relationship between a {\it differential count visibility} $V(u,\varv;q)$ (counts~cm$^{-2}$~s$^{-1}$~keV$^{-1}$), recorded by STIX at count energy $q$, and the corresponding photon images $I(x,y;\epsilon)$ (photons~cm$^{-2}$~s$^{-1}$~keV$^{-1}$~arcsec$^{-2}$) at different photon energy values $\epsilon$ is given by \citep{2007ApJ...665..846P}

\begin{equation}\label{eq:vdef_app_j}
V(u,\varv;q) = \int_X \, \int_Y \, \int_{\epsilon = q}^\infty \, D(q, \epsilon) \, I(x,y;\epsilon) \, e^{2 \pi i(ux + \varv y)} \, d \epsilon \, dx \, dy ~,
\end{equation}
where the spatial integrals extend over the entire field of view of the instrument and $D(q,\epsilon)$ is the (almost diagonal) detector response matrix corresponding to the generation of counts with energy $q$ by photons of energy $\epsilon$.

Combining equations~(\ref{eq:fundamental-1}) and~(\ref{eq:vdef_app_j}) leads to 
\begin{equation}\label{eq:main-result}
\begin{split}
V(u,v;q)  &= \frac{1}{4 \pi R^2}
\, \int_X \, \int_Y \, \int_{\epsilon=q}^\infty \, \int_{E =\epsilon}^{\infty} \mathcal{F}(x,y;E) \, D(q,\epsilon) \, Q(\epsilon, E) \, e^{2 \pi i(ux + \varv y)} \, dE \, d \epsilon \, dx \, dy \\
&= \frac{1}{4 \pi R^2} \int_X \, \int_Y \int_{E = q}^{\infty} \mathcal{F}(x,y;E) \, K(q,E) \, e^{2 \pi i(ux + \varv y)} \, dE \, dx \, dy ~,
\end{split}
\end{equation}
where the \emph{differential count cross-section} $K (q, E)$ (cm$^2$~keV$^{-1}$) is defined as

\begin{equation}\label{eq:ce_cross}
K(q,E) = \int_{\epsilon = q}^E D(q,\epsilon) \, Q(\epsilon, E) \, d \epsilon ~.
\end{equation}
Finally, we introduce the set of \emph{differential electron flux visibilities} (electrons~cm$^{-2}$~s$^{-1}$~keV$^{-1}$)

\begin{equation}\label{eq:evis_app}
W(u,\varv;E) = \int_X \, \int_Y \, \mathcal{F}(x,y; E) \,
e^{2 \pi i(ux + \varv y)} \, dx \, dy ~,
\end{equation}
which correspond to the Fourier transforms of the electron map at the measured spatial frequencies $\{(u_i,\varv_i)\}_{i=1}^{30}$. Combining equations~\eqref{eq:main-result} and~\eqref{eq:evis_app} gives the following relationship between the differential count visibilities $V(u,\varv;q)$ and the corresponding electron flux visibilities:

\begin{equation}\label{eq:result_red}
V(u,\varv;q) = \frac{1}{4 \pi R^2} \, \int_q^{\infty} W(u,\varv;E) \, K(q,E) \, dE ~.
\end{equation}
Equation~\eqref{eq:result_red} is the fundamental result that forms the basis for the reconstruction method introduced in \citep{2007ApJ...665..846P} for use with data from Fourier-transform based imaging instruments such as RHESSI \cite{2002SoPh..210....3L} and STIX \citep{krucker2020spectrometer}. The electron flux visibilities $W(u, \varv;E)$ can be retrieved from the observed count visibilities $V(u,v; q)$ by inverting equation (\ref{eq:result_red}) via any regularization method for spectroscopy \citep[e.g.,][]{1994A&A...288..949P,2003ApJ...595L.127P,2006ApJ...643..523B} and the electron spectral maps for different energies $\mathcal{F}(x,y;E)$ can then be constructed by applying standard image reconstruction methods \citep[e.g.,][]{pianabook} to the set of electron visibilities $\{W(u_i,\varv_i; E)\}_{i=1}^{30}$. We note that as a result of the regularized spectral inversion procedure used to generate the electron visibilities $W(u, \varv; E)$, the electron maps $\mathcal{F} (x, y; E)$ vary smoothly with energy $E$ (unlike the count visibilities $V(u,v; q)$ on which they are based, which suffer from uncorrelated statistical noise in successive count energy channels). In addition, because bremsstrahlung photons at energy $\epsilon$ are produced by electrons at all energies $E > \epsilon$, the electron visibilities $W(u,v;E)$ can be determined up to energies $E$ greater than the maximum count energy observed \citep{2005SoPh..226..317K}. These features of the electron maps are crucial elements in establishing the reliability of the analysis performed in the present study.

\subsection{Electron Acceleration Model}\label{sec:acceleration-model}

We consider a simple acceleration model that invokes a large-scale magnetic-field-aligned electric field ${\cal E}$. Under the combined action of such an electric field and a drag force represented by a collision frequency $\nu$ that is a decreasing function of velocity ($\propto (v_{\rm th}/v)^3$, appropriate to the drag force due to Coulomb collisions with ambient electrons \cite{1962pfig.book.....S}), the equation of motion for a test electron in a target of average density ${\overline n}_{\rm target}$ and temperature $T$ is

\begin{equation}\label{eq:equation_of_motion}
    m_e \, \dot{v} = e \, {\cal E} - m_e \, v \, \nu_C \left ( \frac{v_{\rm th}}{v} \right)^3 = e \, {\cal E} - \frac{4 \pi e^4 \, {\overline n}_{\rm target} \, \ln \Lambda}{2  k_B T}  \, \left ( \frac{v_{\rm th}}{v}  \right)^2\,\,\, ,
\end{equation}
where $e$ (esu) is the electronic charge, $v_{\rm th} = \sqrt{2 k_B \, T/m_e}$ is the thermal speed and the collision frequency for electrons moving at the thermal speed is $\nu_C = 4 \pi  e^4 \, {\overline n}_{\rm target} \, \ln \Lambda/m_e^2 \, v_{\rm th}^3$, with $\ln \Lambda \simeq 25$ the Coulomb logarithm. 
It follows that
\begin{equation}\label{eq:equation_of_motion-dreicer}
    m_e \, \dot{v} = e \left [ {\cal E} - {\cal E}_D \, \left ( \frac{v_{\rm th}}{v}  \right)^2 \right ] \,\,\, ,
\end{equation}
where ${\cal E}_D$ is the Dreicer \cite{1959PhRv..115..238D} field:

\begin{equation}\label{eq:Dreicer Field}
    {\cal E}_D = \frac{2 \pi e^3 \, {\overline n}_{\rm target} \, \ln \Lambda}{k_B T} \simeq 4 \times 10^{-5} \, \frac{({\overline n}_{\rm target}/10^{10} \, {\rm cm}^{-3})}{(T/10^7 \, {\rm K)}} \, \, {\rm V \, cm}^{-1}\,\,\, ,
\end{equation}
and, in the last equality, we have used 1 statvolt $\simeq 300$~V. Equation~\eqref{eq:equation_of_motion-dreicer} shows that for electrons with velocity

\begin{equation}\label{eq:Critical Velocity}
    v > v_{\rm crit} = v_{\rm th} \, \left ( \frac{{\cal E}_D}{{\cal E}} \right )^{1/2} \,\,\, ,
\end{equation}
the force due to the applied field exceeds the collisional drag force; such electrons are therefore accelerated. Furthermore, as they gain speed, the collision frequency and drag force both decrease, resulting in runaway acceleration under the action of the applied field ${\cal E}$. 
The fraction $\eta$ of the ambient population that suffers runaway acceleration is obtained by integrating a Maxwellian distribution over velocities $v > v_{\rm crit}$:

\begin{eqnarray}\label{eq:predicted_nacc_ratio}
   \eta = \frac{n_{\rm acc}}{n_{\rm target}} &=& \frac{\int_{v_{\rm crit}}^\infty e^{-v^2/v_{\rm th}^2} \, dv}{\int_{-\infty}^\infty e^{-v^2/v_{\rm th}^2} \, dv }  = \frac{1}{2} \, {\rm erfc} \left ( \frac{v_{\rm crit}}{v_{\rm th}}\right ) \cr &=& \frac{1}{2} \, {\rm erfc} \left ( \left [ \frac{{\cal E}_D}{{\cal E}} \right ]^{1/2} \right ) \simeq \frac{1}{2 \sqrt{\pi}} \, \sqrt{\frac{{\cal E}}{{\cal E}_D}} \, e^{-{\cal E}_D/{\cal E}} \,\,\, .
\end{eqnarray}
Here ${\rm erfc}(z)$ is the complementary error function and in the last two steps we have used Equation~(\ref{eq:Critical Velocity}) and formula 7.2.14 of \cite{1965hmfw.book.....A}, valid for $v_{\rm crit} \gapprox \, 2 \, v_{\rm th}$ (i.e., ${\cal E} \lapprox 0.25 \, {\cal E}_D$).

\subsection{Collisional Energy Loss Model}\label{sec:energy-loss-model}

We shall initially make the assumption (that will be tested \textit{a posteriori}) that the electron maps inferred from STIX data derive from a cold-target collisional energy loss model, in which electrons are accelerated in a well-defined ``acceleration region'' and lose energy as they stream through the remainder or the flare volume. In a ``cold target'' approximation (i.e., one in which the electrons in question have energies much greater than the mean thermal energy of the electrons in the ambient target), the electrons suffer a systematic energy loss rate with respect to distance $s$ traveled, according to \citep{1972SoPh...26..441B,1978ApJ...224..241E}

\begin{equation}\label{eq:debds}
    \frac{dE}{ds} = - \, \frac{K \, {\overline n}_{\rm target} (s)}{E} \,\,\, ,
\end{equation}
where $K= 2 \pi \, e^4 \ln \Lambda \simeq 2.6 \times 10^{-18} \, {\rm cm}^2 \, {\rm keV}^2$.
The solution of Equation~\eqref{eq:debds} is

\begin{equation}\label{eq:e-N}
E^2(s) = E_o^2 - 2 \, K \, N(s) \equiv E_o^2 - E_{\rm stop}^2(s-s_o) ~,
\end{equation}
where $E_o$ is the electron energy at the injection point $s=s_o$, $N(s) = {\overline n}_{\rm target} \, (s - s_o)$ is the column density (cm$^{-2}$) along the electron trajectory, ${\overline n}_{\rm target}$ is the average value of the target density over the range $[s_o,s]$, and $E_{\rm stop}(s-s_o) = \sqrt{2 \, K \,  {\overline n}_{\rm target} \, (s - s_o)}$ is the initial energy required for an electron to (just) reach position $s$, a distance $(s-s_o)$ from the point of injection. The electron flux continuity equation

\begin{equation}\label{eq:continuity}
    F(E) \ dE = F(E_o) \, dE_o
\end{equation}
\cite[see, e.g.,][with scattering neglected]{1972SoPh...26..441B,1978ApJ...224..241E}, combined with the relation $dE_o/dE = E/E_o$ (from Equation~\eqref{eq:e-N} at fixed $s$), gives, for an accelerated electron spectrum $F(0; E) = A \, E^{-\delta}$,

\begin{equation}\label{eq:collisional-solution}
 F(s; E) = A \, \frac{E}{\left  ( E^2 + E_{\rm stop}^2(s-s_o) \right ) ^{(\delta+1)/2} } \,\,\, .
\end{equation}

Many of the images show evidence of an extended, quasi-homogeneous, region near the top of the loop, suggesting that the acceleration of the energetic electrons extends over a significant volume, so that the point-source expression of Equation~\eqref{eq:collisional-solution} must be generalized to accommodate an extended volume of injection sites $s_o$.  As discussed in \cite{2008ApJ...673..576X,2012ApJ...755...32G,2012A&A...543A..53G}, this can be done in two essential ways:

\begin{enumerate}

\item Assume that the acceleration process dominates over collisional losses within the acceleration region, so that collisional losses only commence once the electrons have left the acceleration region.  Under such an assumption, Equation~\eqref{eq:collisional-solution} generalizes to (for $s > 0$)

\begin{equation}\label{eq:tenuous-acceleration-region}
 F(s; E) = \begin{cases}
 A \, E^{-\delta} \, &: s < s_o \cr
 A \, \frac{E}{\left  ( E^2 + E_{\rm stop}^2(s-s_o) \right ) ^{(\delta+1)/2} } \, &; \, s > s_o\,\,\, .
\end{cases}
\end{equation}

\item Assume that both acceleration and collisional slowing occur simultaneously in the acceleration region. The electron flux spectrum as a function of position is then obtained by convoluting the spectrum \eqref{eq:collisional-solution} with a distribution of accelerated spectra, of (perhaps) spatially-varying amplitude: $F_o(E,s_o) = A(s_o) F_o(E)$, so that \citep[cf. Equation~(19) of][]{2008ApJ...673..576X}

\begin{equation}\label{eq:dense-acceleration-region}
 F(s; E) = \int_{s_o = -\infty}^\infty
 A(s_o) \, \frac{E}{\left  ( E^2 + 2 \, K \, {\overline n}_{\rm target} (\vert s-s_o \vert ) \right ) ^{(\delta+1)/2} } \, ds_o \,\,\, .
\end{equation}
Here we take $A(s_o)$ in the ``square'' form $A(s_o) = A \, (\vert s_o \vert \le L/2)$, leading to

\begin{equation}\label{eq:dense-acceleration-region-square}
 F(s; E) = A \, \, \int_{s_o = -L/2}^{L/2}
  \, \frac{E}{\left  ( E^2 + 2 \, K \, {\overline n}_{\rm target} \vert s-s_o \vert  \right ) ^{(\delta+1)/2} } \, ds_o \,\,\, ,
\end{equation}
which is equation~\eqref{eq:dense-acceleration-region-square-text} in the text.


\end{enumerate}

\bigskip

\section{Data and code availability}
The Interactive Data Language (IDL) code used to reconstruct electron maps and spectra from STIX data is available at \url{https://github.com/theMIDAgroup/STIX_VisibilityInversionSoftware}. The STIX data analyzed in this paper can be downloaded from the STIX Data Center
at \url{https://datacenter.stix.i4ds.net}

\bmhead{Acknowledgements}
We thank Isaiah Beauchamp for stimulating discussions. Solar Orbiter is a space mission of international collaboration between ESA and NASA, operated by ESA. The STIX instrument is an international collaboration between Switzerland, Poland, France, Czech Republic, Germany, Austria, Ireland, and Italy. AV, MP, and AMM are supported by the 'Accordo ASI/INAF Solar Orbiter: Supporto scientifico per la realizzazione degli strumenti Metis, SWA/DPU e STIX nelle Fasi D-E'. AV, AMM, and MP acknowledge the support of the Fondazione Compagnia di San Paolo within the framework of the Artificial Intelligence Call for Proposals, AIxtreme project (ID Rol: 71708). MP acknowledges the support of the PRIN PNRR 2022 Project 'Inverse Problems in the Imaging Sciences (IPIS)' 2022ANC8HL, cup: D53D23005740006. The research by AV, AMM, and MP was performed within the framework of the MIUR Excellence Department Project awarded to Dipartimento di Matematica, Università di Genova, CUP D33C23001110001. AGE was supported by NASA's Heliophysics Supporting Research Program through award 80NSSC24K0244, by the NASA EPSCoR program through award number 80NSSC23M0074 to NASA Kentucky, and by the Kentucky Cabinet for Economic Development.
P M is supported by the Swiss National Science Foundation in the framework of the project Robust Density Models for High Energy Physics and Solar Physics (rodem.ch), CRSII5\_193716.

\bibliography{main}

\end{document}